# A Power Law Approach to Estimating Fake Social Network Accounts

Tushti Rastogi
Oklahoma State University

**ABSTRACT**

This paper presents a method to validate the true patrons of a brand, group, artist or any other entity on the social networking site Twitter. We analyze the trend of total number of tweets, average retweets and total number of followers for various nodes for different social and political backgrounds. We argue that average retweets to follower ratio reveals the overall value of the individual accounts and helps estimate the true to fake account ratio.

## 1. INTRODUCTION

The term Social Network (SN) was coined in 1954 by J.A. Barnes [1]. It represents a social structure of relationships and communication between people, organizations and other entities, on an online site or third party application. It comprises of three main components: nodes, links and communication [2]. A node represents the users or any other entity that has a unique account; links are the interconnections between different nodes; communication represents the interaction among various nodes via these links. The value of any network can be determined by the activity and connectivity among its unique nodes. The nature of the nodes can be: active, dormant, fake or duplicate.

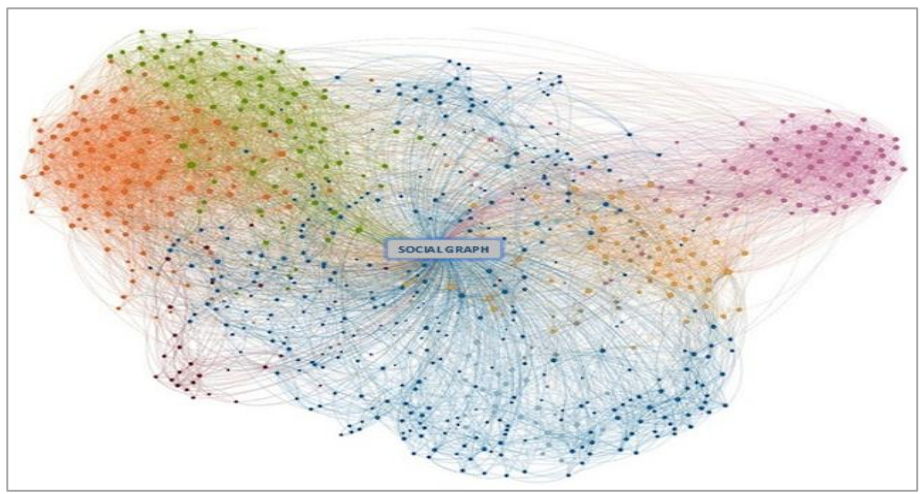

Figure 1. Social Graph depicting nodes and links in LinkedIn [4]

Social Networks require Social Networking Sites (SNSs) which provide a platform to the users to build and expand their social networks. There are two types of SNS: Symmetric and Asymmetric. A Symmetric model makes use of a two-way relationship where both the nodes should have confirmed the relationship to be a member of each other's network e.g. Facebook, LinkedIn, etc. An asymmetric model is how Twitter and Instagram work, where a one-way relationship can be established by the interested party.

Fig. 2 is the social graph generated by Gephi for my personal Facebook account. The different nodes spread out around the node representing my profile at the lower-left end, showcase other Facebook pages and profiles I have visited in the recent past. The software, OutWit Hub version 5.0.0.246 was used to extract data such as source address, destination address and frequency of visiting the same webpage, from my Facebook account. This data was then imported to Gephi, a network analysis tool, which evaluated the various parameters and then provided a graphical representation of the imported data.

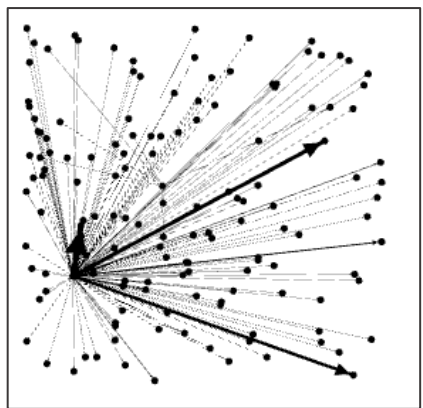

Fig. 2: Social graph created by Gephi for a Facebook account

Today, social networking has become a part and parcel for people of every age group, race, gender and socio-economic background. It has become a vital source for communication, knowledge-sharing, brand publicity and discussions on public issues, news and politics. *"When you walk into a brand, you now ask 'how many followers do they have?' " said Ivan Bart, senior vice president and managing director of IMG Models. – The Wall Street Journal, R.A. Smith, Sep. 3, 2013[3].*

Social Networking tools such as Twitter, Facebook, LinkedIn and the like, play a major role in providing a platform for people to express their ideas and share their experiences. Hence, it lies in the interest of the millions of users around the globe that they get exposed to the true statistics and data which is the foundation of their knowledge and belief.

Social Networks, being a powerful source of information exchange, can pose several challenges and thereby, a need to validate the dynamic nodes and connections. Generating fake followers and likes on social networking sites can be used as an easy technique for entities to gain quick recognition and improve their reputation in the market. Padding the followers to improve the visual identity deludes the naïve audience into thinking that the brand or an artist is genuine and currently trending.

The primary motive behind associating with fake accounts is to attract audiences and customers by targeting the human tendency to: jump on a bandwagon, find an easy way up the ladder of perceived success, have an edge over those in competition and for some it is simply a status concern. These applications have become famous because of our desire to stand out and find short cuts to quick success. Nevertheless, the true patrons are the only ones who would continue to add value by actively participating and getting involved with the brand. They would not get perturbed by the sudden fame or defame of the brand or entity.

In this world of close competition where organizations are constantly competing for the smallest gains, this publicity strategy could easily attract bad press. For example, the recent case of Pepsi and Mercedes-Benz for purchasing fake followers [4], disappointed the patrons of both the brands. An event like this might not have as profound an impact on the entities which already have a strong foundation of patrons as the ones which are battling to flourish for success. However, this does not undermine the need to filter the social networking sites of the redundant nodes. Moreover, with the availability of tools to be able to fake followers on SNSs, there are several prototypes as well as full-fledged online applications available, which could track these spurious accounts.

## 2. BACKGROUND

Social Follow, Twitter Followers Trend, like4like, Magic Liker, Instaliker 1000 and Instamacro are some of the sites and applications which provide fake followers and likes to the users of symmetric networks such as Facebook as well as asymmetric networks such as Twitter and Instagram. A detailed list is provided in Table 1. Some of these sites are free of charge and merely require the users to register with them in order to provide the fake likes (for e.g. like4like) whereas, some of the sites such as buyrealmarketing.com, charge up to $10 per 1000 fake twitter followers.

Table 1. List of sites and applications providing fake followers or likes

| Social Network | Sites/Applications | Activity |
|---|---|---|
| Facebook | LikeFake, like4like, buyrealfbfriends.com | Provides fake likes, friends |
| Twitter | Social Follow, Twitter Followers Trend, granbysportss.org, buyrealmarketing.com | Provides fake likes, followers |
| Instagram | Magic Liker, Instaliker 1000, Famedgram, Monstagram | Provides fake likes, followers |
| | Instamacro | Automates Instagram activity |

These sites make use of software programs such as Twitter Account Creator PRO (shown in Fig.3), which is capable of generating several accounts by sending data through proxy IP addresses and thereby creating fake profiles without actually going through the targeted social networking site [5]. Private proxies i.e. unutilized IP addresses, are used to simulate several users simultaneously. These software creation programs make use of features such as proxy rotation, randomization of page load time, captcha-dodging logic, scraping of profile pictures, etc. in order to avoid detection [6]. Hence, some of the created profiles are a replication of the original one with a few minor changes. Once a fake profile is established, other software programs are used to manage the followers, retweets, likes and messages.

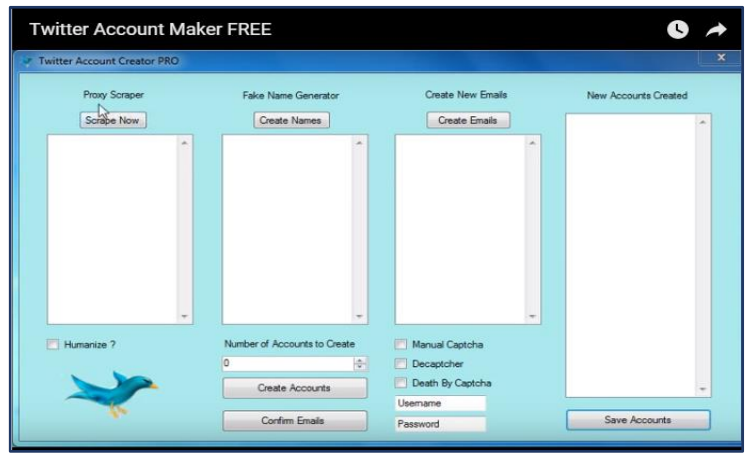

Fig. 3: Twitter Account Creator PRO:
Software that can create thousands of Twitter accounts at a time [5]

With the increase in number of applications providing fake accounts, the availability of tools to detect the fake patrons on social networking sites has also rapidly increased. Some of them are listed in Table 2. These sites make use of the predefined codes and algorithms to verify several criteria such as: sudden spike in the number of followers; followers to followed ratio; no profile picture, description or headline; irrelevant comments; quality of interaction and contribution to pages, diversity of access point (using more than one means to access accounts such as mobile, laptop, etc.); linked accounts; monitoring account activity such as number of Tweets and status updates less than a certain threshold number [5]. For a user account in scrutiny to be genuine, it should at least successfully surpass the cut-off of these validity checks.

Even though checking these criteria can be a good strategy to spot and eliminate fake redundant profiles, they are not sufficient otherwise, the social networking sites would have themselves applied these filters to get rid of all the fake accounts. Moreover, these tools are not as readily available to the public as the fake profile building applications.

Table 2. List of applications available for detecting fake accounts

| Social Network | Sites/Applications |
|---|---|
| Facebook | Social Media Examiner, Social Bakers |
| Twitter | Status People, Social Bakers, Twitter Audit, Twitter Counter, instantcheckmate.com |

Since 2012, many SNSs have been taking measures to purge fake accounts from their sites. For instance, in suspicion of a fake, duplicate or a deceased individual's account, Facebook deleted about 8% of its accounts in March, 2015 [6].

Table 3 presents analysis of the twitter accounts of a few well known entities from different socio-economic backgrounds, generated by Twitter Audit. Twitter Audit analyzes a user's follower list taking into account several characteristics such as number of tweets, ratio of followers to friends, tweet duration, [7] etc. It then evaluates whether the followers are good or bad based on these results. Good represents the number of genuine and active followers whereas, bad represents the number of fake and inactive followers.

Table 3. Analysis of few Twitter accounts using Twitter Audit [7]

| S.No | Name | Designation | Total Followers | Good % | Fake % | Good | Fake | Audit date |
|---|---|---|---|---|---|---|---|---|
| 1 | Dick Costello | CEO of Twitter | 1.63M | 76 | 24 | 1,229,531 | 392,541 | 2013 |
| 2 | Jack Dorsey | Co-Founder of Twitter | 3.45M | 77 | 23 | 2,661,944 | 781,709 | 2013 |
| 3 | Bill Gates | Co-Founde of Microsoft | 27.4M | 33 | 67 | 9,042,000 | 18,358,000 | 2014 |
| 4 | Barack Obama | President | 70.3M | 37 | 63 | 26,063,101 | 44,187,845 | 2015 |
| 5 | Sunidhi Chauhan | Indian Singer | 2.4M | 36 | 64 | 500,607 | 889,969 | 2014 |
| 6 | Priyanka Chopra | Indian Actress | 12.9M | 23 | 77 | 879,642 | 2,961,590 | 2013 |
| 7 | Narendra Modi | PM of Delhi, India | 18.2M | 32 | 68 | 4,940,844 | 10,499,293 | 2013 |
| 8 | AR Rehman | Music Composer | 10.3M | 24 | 76 | 635,933 | 1,959,711 | 2014 |
| 9 | Sachin Tendulkar | Indian Cricketer | 9.66M | 20 | 80 | 705,149 | 2,768,489 | 2013 |
| 10 | Prannoy Roy | Co-Founder & Executive Co-Chairperson of NDTV | 0.46M | 54 | 46 | 180,728 | 153,954 | 2015 |
| 11 | Subhash Kak | Professor | 1.06K | 92 | 8 | 974 | 86 | 2016 |

Barack Obama, is at the top of the list with the highest number of followers but only 36% of his followers are real. This might be because it is his last year of presidency and so the focus has been shifted more towards the current presidential candidates. On the other hand, OSU professor, Dr. S. Kak, despite having the lowest number of followers, display the highest percentage of real followers. Therefore, probability of fake nodes associated with social accounts is independent of their total followers. In general, this method of evaluation of an individual's twitter account helps us determine the quality of the social network.

## 3. GENERAL PROPERTIES OF SOCIAL NETWORKS

Small-world networks and Scale-free networks are the two main properties of the social networks. In a *small-world network*, all nodes are not directly linked to each other but most of them can be interconnected by taking a few extra steps, exhibiting a short average path length [8]. Bridges are present to shorten the average distance between the clusters of locally connected nodes. Due to its transitivity and ability of weak links to connect across clusters [9], Small-world model is cost-effective and supports complexity. Hence, it finds its use in many areas such as social networks, internet architecture and even in cognition networks.

Mathematically, a small-world network can be represented by the following equation:

$$L \propto \log N \qquad (1)$$

L is the distance between any two random nodes and N is the total number of nodes in the network [8], [15]. This equation represents that the distance between any two random nodes is proportional to the log of network size. The following figures represents a small-world network.

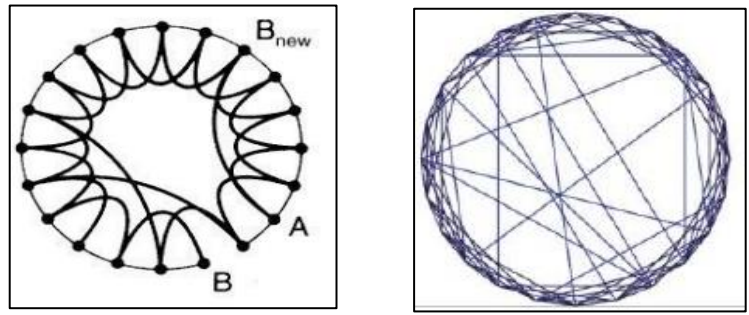

Fig. 4: Small-world network [10], [12]

*Scale-free networks* on the other hand follow the power law distribution showcasing a skewed distribution of links. It can be constructed by adding more nodes to the existing network and forming links with the existing nodes in a probabilistic manner such that it results in a hierarchical structure [12]. The following figures represent a scale-free network with grey circles as Hubs:

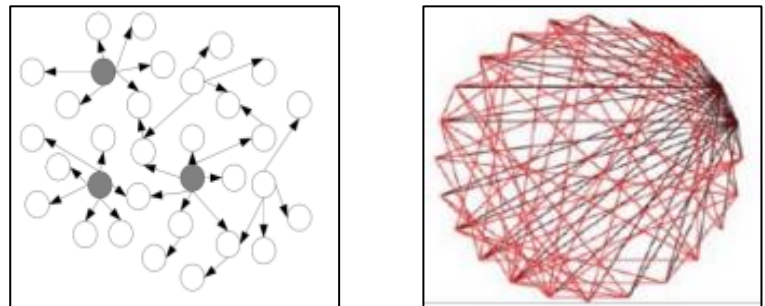

Fig. 5: Scale-free network [11], [12]

Mathematically, the number of links (k) originating from any given node following the power law can be represented as follows:

$$P(k) \sim k^{-\gamma} \qquad (2)$$

Typically, value of γ lies between 2 and 3 (2 < γ < 3) [13]. The nodes with highest degree of links are referred to as Hubs. These hubs are further connected with relatively lower degree nodes. Scale-free model offers good fault tolerance as failure of small degree nodes would not affect the hubs and in the event of hub-failure, other hubs will maintain the connectivity [13].

## 4. POWER LAW

Power law gives the relationship between two quantities where, a relative change in one quantity is reflected as a proportional relative change in the other quantity, regardless of the initial value of both the quantities. Mathematically, it can be represented as:

$$y = ax^k \qquad (3)$$

Taking Log on both sides, $\quad \text{Log}(y) = \text{Log}(a x^k)$

$$\text{Log}(y) = \log(a) + k \log(x) \qquad (4)$$

where 'x' and 'y' are the variables of interest, 'k' is the power law exponent and 'a' is a constant. For increasing or decreasing functions, k is positive or negative respectively. As seen in equation (4), power law adopts a linear relationship if the variables are plotted on a logarithmic scale. Power law is frequently used to determine the underlying properties of social, scientific, human as well as natural systems.

The power law can be used to reveal the characteristics of a social network. As the network evolves with time, large number of new edges might get added to nodes which already have a large number of links, thereby increasing the degree of nodes disproportionately. This results in a few highly connected nodes and many weekly connected nodes displaying a long-tailed degree distribution of a scale-free network [9].

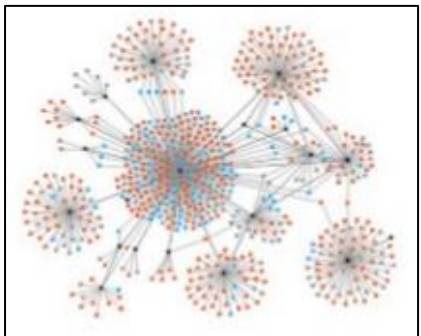 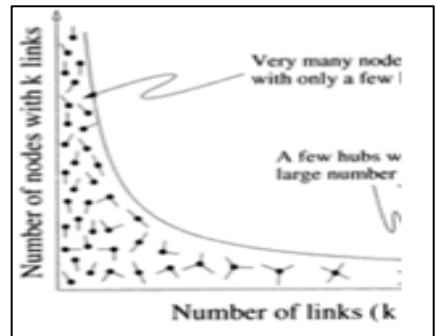

Fig. 6: Network Evolution [9]      Fig. 7: Scale-free network-power law [14]

## 5. ZIPF'S LAW

Zipf's law is one of most useful power-law distributions in the field of social and physical data approximation. It states that the frequency of occurrence of an element in a well-defined sample space is inversely proportional to its rank in that space. For instance, the frequency of utterance of a word in any natural language is inversely proportional to its rank in the frequency table [16], [24]. Hence, the frequency of occurrence of the second most frequent word will be half the frequency of the most frequent word; the frequency of the third most frequent word will be one-third of the most frequent word and so on. Hence, the following relation can be established [19]:

$$X = \frac{F}{\sum_{n=1}^{N} n} \qquad (5)$$

Where, X is the frequency of occurrence of the element at rank 'n' and F is the frequency of the element at rank 1. Equation (5) illustrates that the value of an element at rank n (where, n∈N) will be $1/n^{th}$ times the value of the element at rank one.

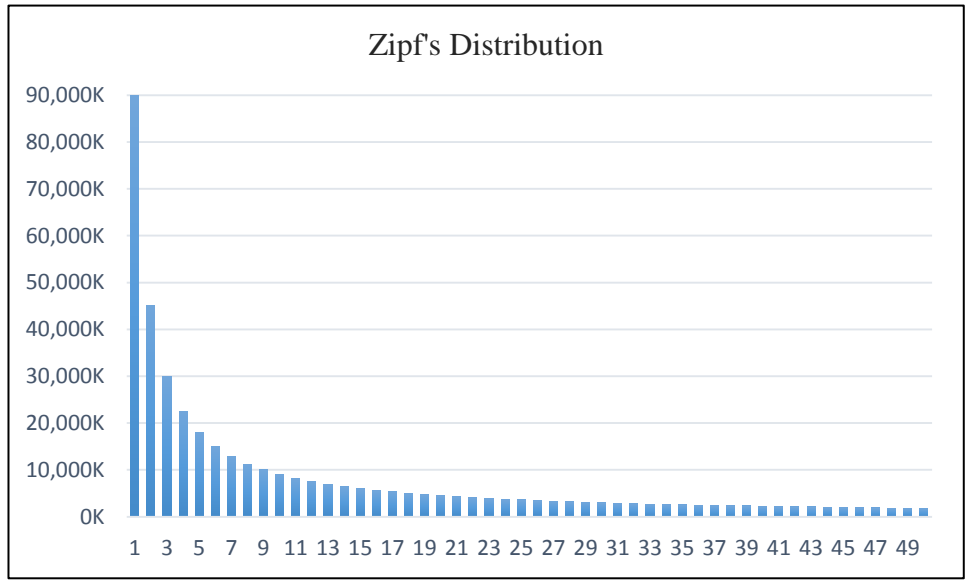

Fig. 8: Zipf's Distribution

A random data set of 50 points starting at 90 million was generated to graphically represent the Zipf's distribution. Rank is plotted on the abscissa and the value of each rank is plotted on the ordinate. As seen in Fig. 8, the resulting Zipf's distribution follows a power-series model defined by equation (2). Rank 1 is twice as large as rank 2, thrice as large as rank 3 and 50 times greater than the last data point with rank 50.

Zipf's law is a good method to estimate the value of an element if the number of elements is a random variable following power-law distribution where, the specific exponent value in the power law characterizes its distribution [24]. Determining cumulative distribution function and probability mass function for Zipf's law is a good method for studying the characteristics of data [16].

# 6. DATA ANALYSIS

## 6.1 ANALYSIS OF OVERALL DATA

Overall Analysis of Total Tweets, Average Retweets and Total Followers:

Table 4. Data from Twitter for the top 12 ranks

| Rank | Total Tweets | Average Retweets | Total Followers |
|---|---|---|---|
| 1 | 96.20K | 51.70K | 84.80M |
| 2 | 54.20K | 41.40K | 77.50M |
| 3 | 48.50K | 24.20K | 73.30M |
| 4 | 41.70K | 20.71K | 71.00M |
| 5 | 31.40K | 14.80K | 41.50M |
| 6 | 30.70K | 9.32K | 40.60M |
| 7 | 27.30K | 5.93K | 35.10M |
| 8 | 23.70K | 4.94K | 27.80M |
| 9 | 23.60K | 4.93K | 27.80M |
| 10 | 21.10K | 4.70K | 21.40M |
| 11 | 21.00K | 4.01K | 20.40M |
| 12 | 18.40K | 3.73K | 19.90M |

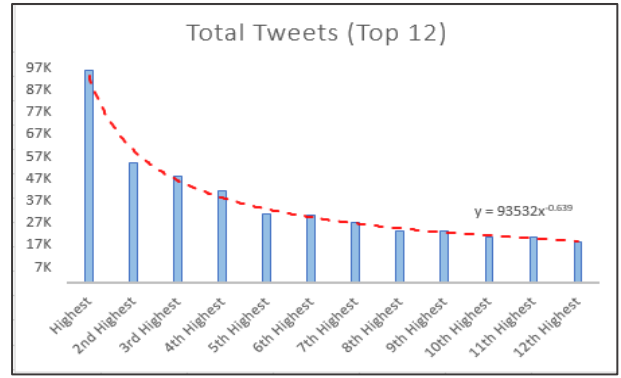

Fig. 9.1: Analysis of Total Tweets

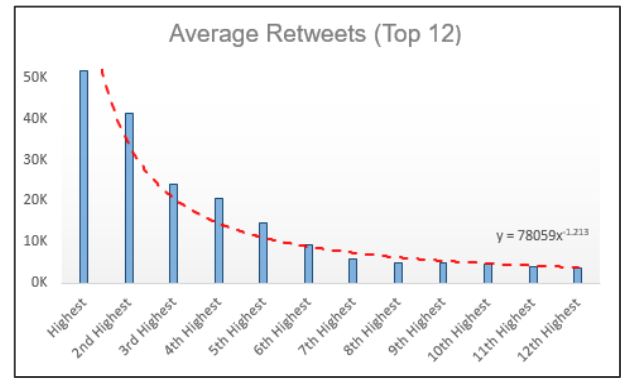

Fig. 9.2: Analysis of Average Retweets

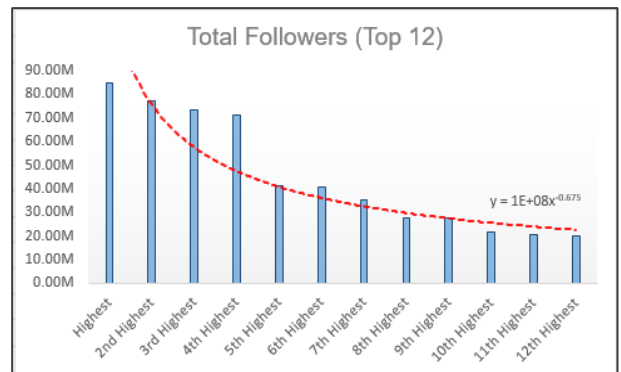

Fig. 9.3: Analysis of Total Followers

Table 4 presents the total tweets, average retweets and total followers of the top 12 ranks (evaluated as mutually exclusive parameters) from the Twitter data (appendix) put together for the 70 individuals.

Graphical representation is used to further analyze the output of the data for Zipf's distribution. The parameter values are plotted on the abscissa and their respective rank is plotted on the ordinate. A dotted line running across the graphs represents the power series model. As we can see from Figures 9.1, 9.2 and 9.3, the trend line closely follows the power law for total tweets and average retweets showcasing a good fit for Zipf's distribution. In the Total Followers case, it may not be a good fit for the upper tail but, 5$^{th}$ rank onwards, the lower tail gets closer to the power series' trend line.

Hence, the data arranged in descending order is observed to follow Zipf's law for all the three cases.

## 6.2 ANALYSIS OF CELEBRITIES (Movies and Music industry)

Table 5. Data retrieved from Twitter for Celebs

| Rank | Total Tweets | Average Retweets | Total Followers |
|---|---|---|---|
| 1 | 54.20K | 51.70K | 84.80M |
| 2 | 48.50K | 41.40K | 77.50M |
| 3 | 30.70K | 24.20K | 73.30M |
| 4 | 27.30K | 20.71K | 40.60M |
| 5 | 23.70K | 9.32K | 35.10M |
| 6 | 23.60K | 5.93K | 27.80M |
| 7 | 18.40K | 4.93K | 21.40M |
| 8 | 17.60K | 4.70K | 19.90M |
| 9 | 16.90K | 4.01K | 18.40M |
| 10 | 16.90K | 3.73K | 16.90M |
| 11 | 11.90K | 3.25K | 16.40M |
| 12 | 8.40K | 2.00K | 15.20M |
| 13 | 8.18K | 1.69K | 14.90M |
| 14 | 8.10K | 1.21K | 13.60M |
| 15 | 6.90K | 1.02K | 13.10M |
| 16 | 5.60K | 0.77K | 11.60M |
| 17 | 4.43K | 0.71K | 11.60M |
| 18 | 4.39K | 0.67K | 10.50M |
| 19 | 4.16K | 0.62K | 8.83M |
| 20 | 4.10K | 0.54K | 6.94M |
| 21 | 4.10K | 0.46K | 6.40M |
| 22 | 3.76K | 0.43K | 6.32M |
| 23 | 2.44K | 0.42K | 6.24M |
| 24 | 1.36K | 0.34K | 4.92M |
| 25 | 1.25K | 0.13K | 4.54M |
| 26 | 1.05K | 0.10K | 4.04M |
| 27 | 0.70K | 0.06K | 2.91M |
| 28 | 0.62K | 0.06K | 2.47M |
| 29 | 0.41K | 0.05K | 2.08M |
| 30 | 0.40K | 0.04K | 1.47M |
| 31 | 0.37K | 0.02K | 0.11M |

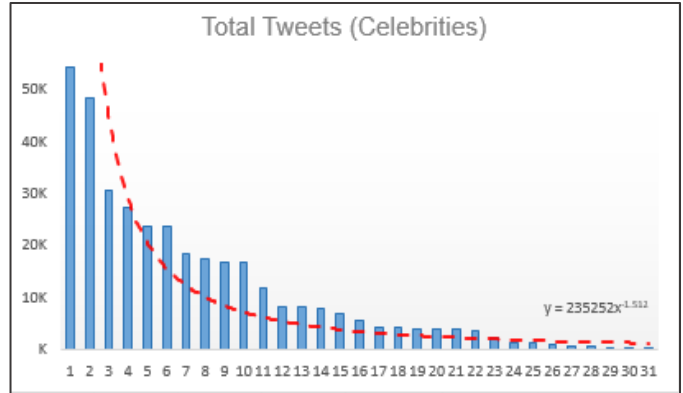

Fig. 10.1: Analysis of Total Tweets for Celebs

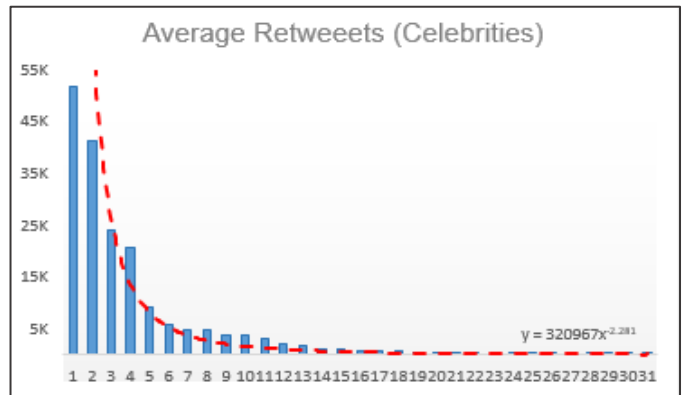

Fig. 10.2: Analysis of Average Retweets for Celebs

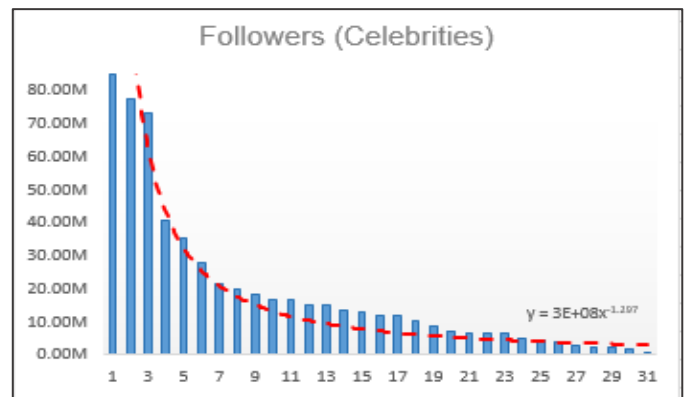

Fig. 10.3: Analysis of Total Followers for Celebs

The above figures represent the trend of celebrity Twitter accounts on the basis of total tweets, average retweets and total number of followers they have. It can be inferred that all the three parameters conform to the Zipf's law and show a decreasing trend proportional to $1/n^{th}$ of the highest rank where, n is the position of the rank.

## 6.3 ANALYSIS OF POLITICIANS

Table 6. Twitter data for Politicians

| Rank | Total Tweets | Average Retweets | Total Followers |
|------|--------------|------------------|-----------------|
| 1 | 31.40K | 4.94K | 71.00M |
| 2 | 15.20K | 2.28K | 18.50M |
| 3 | 14.70K | 2.26K | 7.16M |
| 4 | 13.90K | 1.39K | 5.73M |
| 5 | 10.70K | 1.37K | 3.89M |
| 6 | 7.86K | 1.08K | 1.71M |
| 7 | 4.78K | 0.68K | 1.39M |
| 8 | 3.18K | 0.62K | 0.95M |
| 9 | 2.20K | 0.16K | 0.25M |

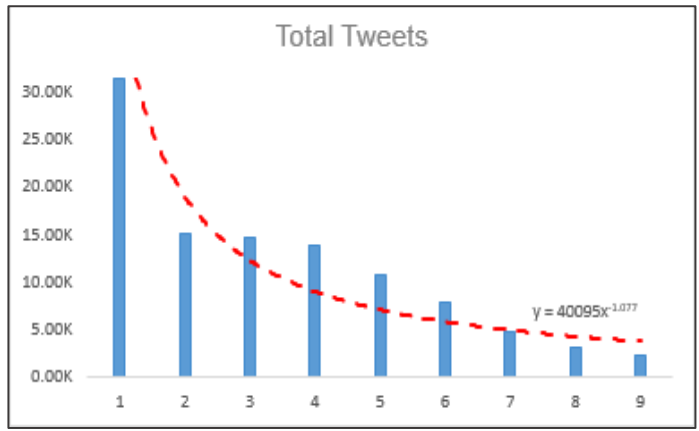

Fig. 11.1: Analysis of Total Tweets for Politicians

Table 6 presents the Twitter data retrieved for current American presidential candidates (such as Donald Trump and Hillary Clinton), and other politicians such as an Indian politician, Narendra Modi and a British politician, David Cameron.

In the Politicians' subset, graphs for all the three parameters (i.e. 11.1, 11.2 and 11.3) prove to be a good fit for the Zipf's distribution. However, the graph representing the distribution of total followers is the best of them all.

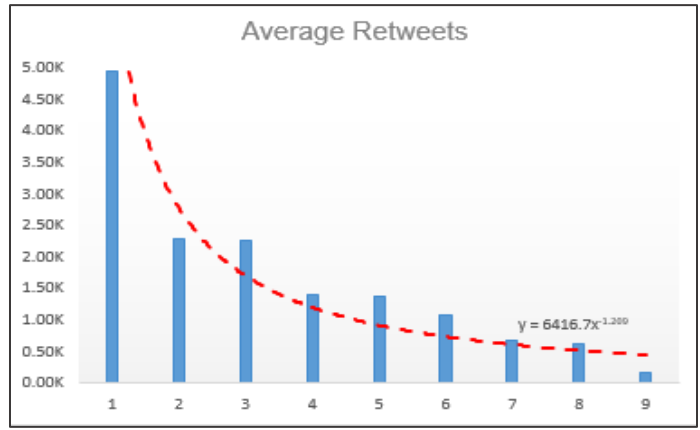

Fig. 11.2: Analysis of Average Retweets for Politicians

Hence, the total number of followers of the politician at rank 2 is approximately half of the total number of followers of Barack Obama who is at rank 1 in case of the total follower parameter. Similarly, the politician who has the 9$^{th}$ largest size of followers will be sitting at rank 9.

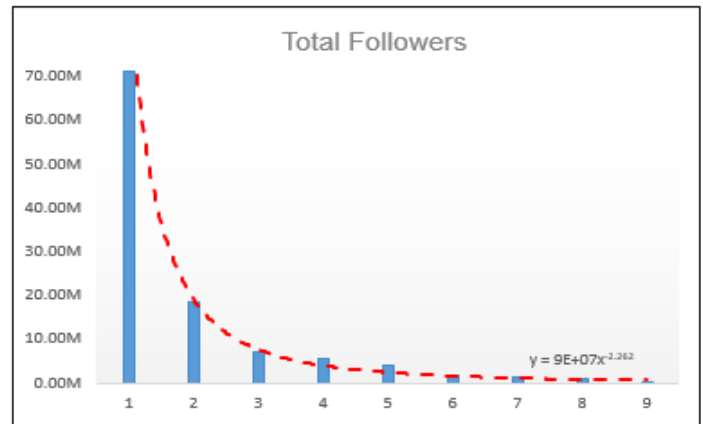

Fig. 11.3: Analysis of Total Followers for Politicians

## 6.4 ANALYSIS OF SPORTSMEN

Table 7. Twitter data for Sportsmen

| Rank | Total Tweets | Average Retweets | Total Followers |
|------|--------------|------------------|-----------------|
| 1 | 96.20K | 3.23K | 9.99M |
| 2 | 21.00K | 2.82K | 9.23M |
| 3 | 13.30K | 0.63K | 6.10M |
| 4 | 4.20K | 0.54K | 5.10M |
| 5 | 2.70K | 0.22K | 3.59M |
| 6 | 2.10K | 0.14K | 3.54M |
| 7 | 0.78K | 0.08K | 2.07M |
| 8 | 0.78K | 0.07K | 1.69M |
| 9 | 0.42K | 0.02K | 1.60M |

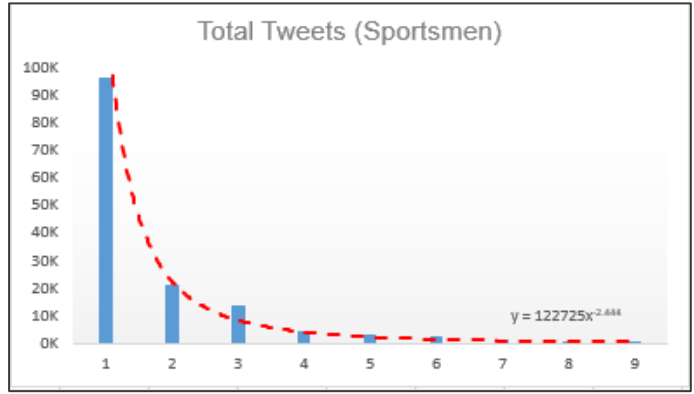

Fig. 12.1: Analysis of Total Tweets for Sportsmen

Table 7 presents the total number of tweets, average tweets and total number of followers for the following 9 sports people: Amir Khan, Mahendra Singh Dhoni, Sachin Tendulkar, Chad Johnson, Wasim Akram, Michael Clarke, Sania Mirza, Rafael Nadal and Serena Williams.

Indian cricketer, Sachin Tendulkar has the maximum number of followers and highest average retweets placing him at rank 1 and the famous Spanish tennis player, Rafael Nadal at rank 2.

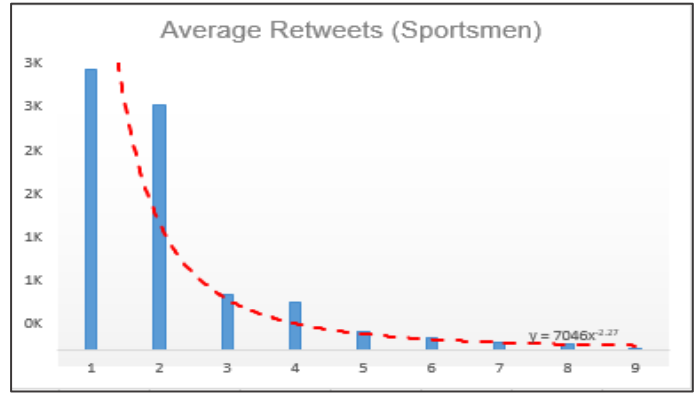

Fig. 12.2: Analysis of Total Tweets for Sportsmen

Power law trend line is represented by a dashed line running along the data output in the graphs. All the three graphs are a good fit for the power law in speculation, with the graph representing total tweets as the best of them all.

In case of sportsmen, power law equations have a greater negative exponent as sportsmen don't have as many active followers as celebrities and politicians do. Hence, the deviation in adjacent node values is greater.

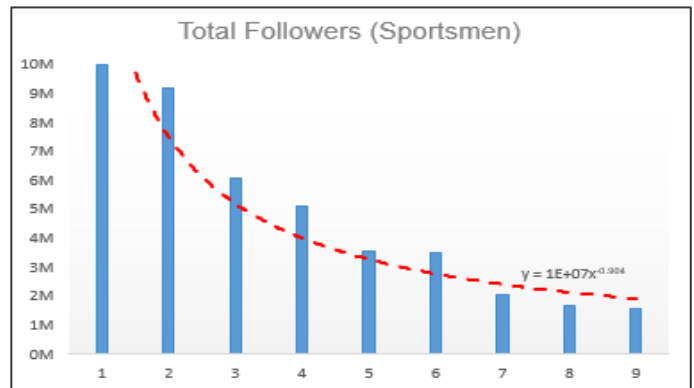

Fig. 12.3: Analysis of Total Tweets for Sportsmen

## 7. POWER LAW HYPOTHESIS

Total tweets and follower data was collected from the authentic twitter accounts of all the 70 individuals. Average retweets were evaluated by taking into account retweets of the top fifteen posts. P is the ratio of average retweets to total number of followers as shown in equation (5). The ratio 'P' is then normalized such that it returns a whole number on taking the logarithmic value. Log (N) values are then plotted on the x-axis of the graph ranging from 1 to 4 units. The abscissa is further divided into bins of 0.2 units to evaluate the number of nodes falling in each bin.

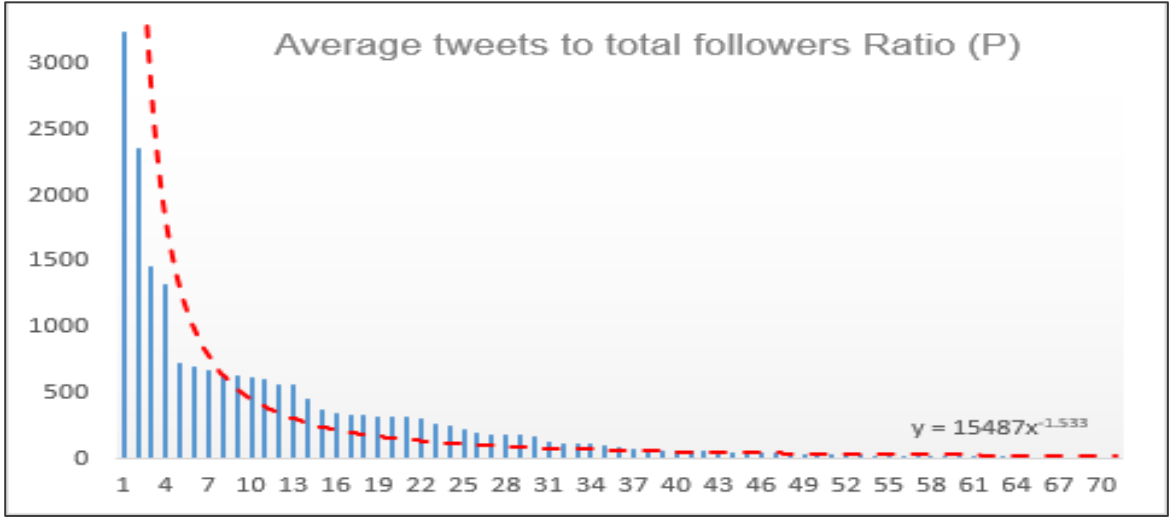

Fig. 13. Analysis of the parameter 'P'

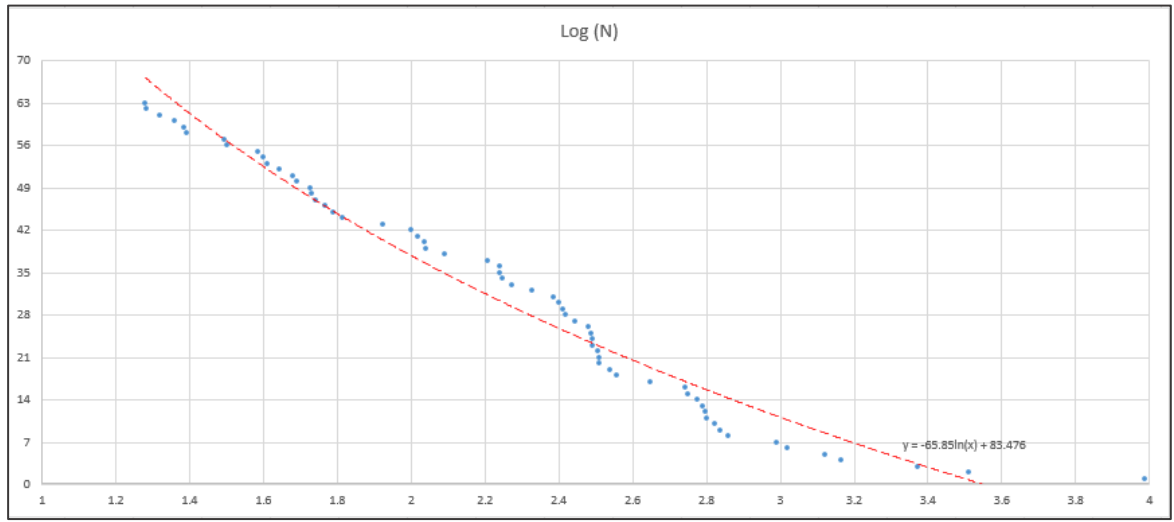

Fig. 14. Log (N) Analysis

Fig. 13 shows that the ratio P also follows the power series model. Since logarithm is an inverse operation of exponentiation, Log N values, as depicted in Fig. 14, form a linear plot. Thus, the power-law holds true for our data set.

Table 8. Bin density

| Bin | Number of Nodes |
|---|---|
| 0.5 - 1.0 | 2 |
| 1.0 - 1.2 | 4 |
| 1.2 - 1.4 | 7 |
| 1.4 - 1.6 | 4 |
| 1.6 - 1.8 | 9 |
| 1.8 - 2.0 | 3 |
| 2.0 - 2.2 | 4 |
| 2.2 - 2.4 | 7 |
| 2.4 - 2.6 | 13 |
| 2.6 - 2.8 | 6 |
| 2.8 - 3.0 | 5 |
| 3.0 - 3.2 | 3 |
| 3.2 - 3.4 | 1 |
| 3.4 - 3.6 | 1 |
| 3.6 - 3.8 | 0 |
| 3.8 - 4.0 | 1 |

$$P = \frac{Average\ Retweets}{Total\ Followers} \quad (6)$$

$$N = P * 10^6 \quad (7)$$

The dashed line in Fig. 12 represents the logarithmic trend line. Log N values are a good fit for the trend line with the following equation:

$$y = -65.85\ln(x) + 83.476$$

Table 8 gives the number of individuals or nodes falling in different bins. The node density increases at first, reaches the maximum and then gradually decreases. Largest number of nodes fall in the bin ranging from 2.4 – 2.6.

Even though the distribution of nodes is not uniform across the various bins, they display linear characteristics when plotted together.

**POWER-LAW EQUATIONS**

Table 9 provides the power-law equations of the three parameters used for analysis of the overall data and its subsets. This data can be used for value-analysis of the three classes: celebrities, politicians and sportsmen. The exponential values are very different for all the three classes as their pool of interested audience differs. In case of sportsmen, the exponents have a greater negative value depicting that there is a quick fall in the value of adjacent elements. This might be because people are more drawn towards politicians and the glitz and glamour of celebrities than sportsmen.

Table 9. Power-Law Equations

| Power Law Equations of Best Fit | | | |
|---|---|---|---|
| | Total Tweets | Average Retweets | Total Followers |
| Overall data | $62750\ x^{-0.915}$ | $76853\ x^{-1.197}$ | $10^7\ x^{-0.62}$ |
| Celebrities | $235252\ x^{-1.512}$ | $320967\ x^{-2.281}$ | $3*10^7\ x^{-1.297}$ |
| Politicians | $40095\ x^{-1.077}$ | $6416.7\ x^{-1.209}$ | $9*10^6\ x^{-2.262}$ |
| Sportsmen | $122725\ x^{-2.444}$ | $7046\ x^{-2.27}$ | $10^6\ x^{-0.904}$ |

## 8. CONCLUSION

The underlying nature of the social network system is captured by the power law. Even though the total tweets, average retweets, total followers and the ratio of average retweets to total followers are unrelated parameters, they all follow the crowded upper-tail and sparsely populated lower-tail characteristics of Zipf's power-distribution, further demonstrating the network evolution.

The parameters associated with the power law vary amongst different category of individuals. This can help check on whether the number of followers in a particular category are real or fake.

The evaluation of the ratio P proved to be a good measure for analysis of the individual nodes and their cumulative contribution to the social network. A higher value of P is desirable as it signifies the availability of a large number of active nodes involved with the target node, thereby, improving the quality of the overall network. It further provides a platform for the comparison of connectivity among the various dynamic and complex nodes available within a particular network.

The power law equations provided in Table 9 reveal that, greater the negative exponential value, greater is the drop in values of the adjacent elements and smaller is the size of the interested parties for that particular category. The exponential value plays a major role in determining the characteristics of the distribution of elements in any system. Hence, it can be very useful in revealing the nature of the network, co-relation with other networks and systems, the strengths and weaknesses of a network, and most importantly the network traffic.

## REFERENCES

[1] J.A. Barnes, Class and Committees in a Norwegian Island Parish, Human Relations, 7 pp. 39-58, 1954.
[2] S. Yu and S. Kak, Social Network Dynamics: An Attention Economics Perspective. In Social Networks: A Framework of Computational Intelligence, Edited by Witold Pedrycz and Shyi-Ming Chen. Springer, 2014.
[3] R.A. Smith, As Fashion Week Approaches, Runway Models Strut on Social Media, The Wall Street Journal, Sep 3, 2013.
[4] S.F.J. Mak, Research in Connectivism, Learner Weblog, Nov 10, 2012.
[5] J. Dougherty, How Fake Twitter Follower Bots Work, Leaders West-Digital Marketing Journal, 2012.
[6] J. Parsons, Facebook's war continues against fake profiles and bots, Huffpost Business, May 22, 2015
[7] Data collected from Twitter Audit.
[8] http://www.nature.com/nature/journal/v393/n6684/full/393440a0.html
[9] G. Celiotis, Social Network Analysis, CNM Social Media Module, Slideshare, Feb 25, 2015

[10] L. Adamic, Small World Networks ppt., at Brynmawr.edu, adapted from slides by Lada Adamic, UMichigan, 2013
[11] R.B. Yates, C. Castillo, V. Lopez, Characteristics of the web of Spain, Volume 9, 2015
[12] J. Lu, Brain Networks, Technology, Brain-informaticsweco-lab, Slideshare, Feb 22, 2010.
[13] A. Clauset, C. R. Shalizi, M. E. J. Newman, Power-Law Distributions in Empirical Data, 2009. SIAM Review 51 (4): 661–703. ArXiv: 0706.1062.
[14] J. Hagel III, The Power of Power Laws, Edge Perspectives with John Hagel, May, 2007.
[15] L. Beauguitte, C. Ducruet, Scale-free and small-world networks in geographical research: A critical examination, halshs-00623927, Sep 15, 2011
[16] S. Fagan, R. Gençay, An introduction to textual econometrics, in Ullah, Aman; Giles, David E. A., Handbook of Empirical Economics and Finance, CRC Press, 2010, pp. 133–153.
[17] S. Soh, G. Lin, and S. Kak, Connectivity in Social Networks, arXiv:1506.03377,2015
[18] C. Lieber, The Dirty Business of Buying Instagram Followers, racked.com, Sep 11, 2014.
[19] I. Moreno-Sánchez, F. Font-Clos, A. Corral, Large-Scale Analysis of Zipf's Law in English Texts, 2016. PLoS ONE. doi:10.1371/journal.pone.0147073.
[20] "How to identify fake likes and followers" at socialdraft.com, May 2, 2015.
[21] S. Kak, Friendship Paradox and Attention Economics, arXiv:1412.7049
[22] S.A. Pilgaonkar, A Study of Network Paradoxes, Master's Thesis, OSU, Stillwater, 2014.
[23] J. P. Onnela et al. Structure and tie strengths in mobile communication networks, Proc. of the National Academy of Sciences 104 (18): 7332-7336, 2007. arXiv: physics/0610104
[24] S. Kak, Two power series models of self-similarity in social networks. arXiv:1506.07497